# Near-field light-bending photonic switch: physics of switching based on three-dimensional Poynting vector analysis

Liyang Yue[1,*], Zengbo Wang[1], Bing Yan[1], Yao Xie[2], Y. E. Geints[3], Oleg V. Minin[4], and Igor V. Minin[4]

[1]School of Computer Science and Electronic Engineering, Bangor University, Dean Street, Bangor, Gwynedd, LL57 1UT, UK
[2]School of Physics and Optoelectronic Engineering, Xidian University, No. 2 South Taibai Road, Xi'an, Shaanxi, 710071, China
[3]V. E. Zuev Institute of Atmospheric Optics, Siberian Branch, Russian Academy of Sciences, 1 Zuev square, Tomsk, 634021, Russia
[4]Tomsk Polytechnic University, 36 Lenin Avenue, Tomsk, 634050, Russia
Email: l.yue@bangor.ac.uk

**Abstract**
Photonic hook is a high-intensity bent light focus with a proportional curvature to the wavelength of the incident light. Based on this unique light-bending phenomenon, a novel near-field photonic switch by means of a right-trapezoid dielectric Janus particle-lens embedded in the core of a planar waveguide is proposed for switching the photonic signals at two common optical communication wavelengths – 1310 nm and 1550 nm by using numerical simulations. The signals at these two wavelengths can be guided to different routes according to their oppositely bent photonic hooks to realise wavelength selective switching. The switching mechanism is analysed by an in-house developed three-dimensional (3D) Poynting vector visualisation technology. It demonstrates that the 3D distribution and number of Poynting vector vortexes produced by the particle highly affect the shapes and bending directions of the photonic hooks causing the near-field switching, and multiple independent high-magnitude areas matched by the regional Poynting vector streamlines can form these photonic hooks. The corresponding mechanism can only be represented by 3D Poynting vector distributions and is being reported for the first time.

**Keywords**: photonic hook, optical switch, Poynting vector, vortex.

## 1. Introduction

Switch is the most basic component to make, break or divert the connection from one conductor to another in a circuit. The conventional electronic switch is a solid-state mechanical device though the sensors can be used to automate it in certain working environments. The concept of the switch can be used for signal processing and other applications of modern communication technologies, and activation/deactivation of signals and data routing are the major functions of a switch, sometimes called logic signal switch, in here [1]. With the development of data technology and mobile computing, the optical signal in the form of light replaced the electronic signal to become the mainstream of data transmission with the aid of optical fibres [2, 3]. For this reason, photonic switch is developed to control the light (optical signals) propagation with high efficiency, and its switching time is much faster than that of a conventional electronic switch due to the unparalleled speed of light [4, 5]. In this field, wavelength switching is a technology used in optical communication to select individual wavelengths of light and forward the selected input light to the separate paths for specific data routing [6]. Meanwhile, many optical fibres are optimised for the optical signals at the wavelengths of 1310 nm and 1550 nm because of their lower losses in the glass fibre [7]. Wavelength division multiplexing (WDM) and Dense WDM (DWDM) are the technologies that enable a number of separate optical signals with different wavelengths to be transmitted in an optical fibre [8, 9], but the sizes of these devices limit their applications for the future photonic integrated circuits and lab-on-a-chip devices. As an alternative, planar dielectric waveguide in the form of a dielectric film (core) sandwiched between cladding layers better applies to these integrated optoelectronic devices because of a relatively simple and compact structure and the consequent production advantages [10, 11]. Theoretically, the optical signal can be fast and accurately transmitted in the core of a planar waveguide relying on the total internal reflection occurring at the interface between the core and cladding [12].

In this paper, a novel all-dielectric photonic switch is realised by the near-field light-bending effect of photonic hook which is a curved and localised high-intensity light beam focused by a dielectric microparticle [13]. It was the only other instance of artificially bent light apart from Airy beam when it was invented [13, 14]. Meanwhile, it is highly related to the concept of photonic jet or can be directly understood as a photonic jet asymmetrically shifting in optical phase. Typically, it is produced by the light focusing through a dielectric Janus microparticle with the asymmetry in geometrical shape or internal refractive index [14-16]. This wavefront curvature normally appears in the shadow direction of the particle and significantly influence the electric field distribution in that area. Minin et al. experimentally generated a photonic hook in terahertz (THz) band for the first time using a relatively simple experimental setup, meanwhile proved the proportional relationship between the wavelength of incident light and curvature of photonic hook [15]. Based on this finding, Geints et al. raised the proofs-of-concept of similar photonic switches, and herein photonic hooks could function like switching channels after passing several differently configured dielectric particles, such as the flipped prism, Janus bar, and off-axis Fresnel zone plate [17]. Besides, most photonic hooks are graphically represented by the distributions of electric field and Poynting vector on the median section. This two-dimensional (2D) plot is sufficient to characterise the complete shape and curvature of a photonic hook, however, it is difficult to analyse the light-bending process and the physics behind the phenomenon because the influence of the photonic energy that flows into the median section from other planes cannot be quantified, even visualised [18]. Yue et al. developed a 3D mapping technology to track the power flow (Poynting vectors) of a photonic jet focused by a spherical mesoscale micro-particle, and the circulation and convergence of 3D Poynting vectors inside the dielectric particle were demonstrated for the first time [19]. The same technology is used in this study to solve the above issue.

In this article, we propose a novel photonic hook light-bending optical switch by means of a conventional right-trapezoid-shaped dielectric particle embedded in the core of a planar waveguide. The operation of signal switching at two communication wavelengths – $\lambda$ = 1310 and 1550 nm are numerically simulated. Its setup and operating wavelengths are designed for the applications of optical communication and are different from those considered in [17]. The separation of switching channels for these two wavelengths is shown in the overall distribution of the electric field and profiles of the generated photonic hooks. The corresponding simulation data is analysed by 3D Poynting vector visualisation technology to deeply investigate the physics of switching and its mechanism.

## 2. Results and discussion
### 2.1 Concept and modelling
The numerical model of the proposed photonic switch is built in the commercial finite integral technique (FIT) software package – CST Microwave Studio (CST). A right-trapezoid-shaped particle with the refractive index of $n$ = 2 made of high-index dielectric materials, e.g. Boron nitride (BN), MICROPOSIT® S1800 photoresist, etc., is implanted in the facet of the quartz core of a planar waveguide with $n$ = 1.44 [20-22]. The assumed planar waveguide is able to ideally approximate the plane wave propagation and eliminate the interference for the transmitted light at the wavelengths of 1310 nm and 1550 nm and the implanted particle with a small relative dimension to the core ($d/D \lesssim$ 0.08, where $d$ and $D$ are the particle size and thickness of the core of the planar waveguide, respectively) [23, 24]. The sketch of this design is shown in Figure 1 (a) I, and the diagrams of Figure 1 (a) II and III illustrate the effect of a trapezoid particle switching the incident lights with short and long wavelengths which are 1310 nm and 1550 nm in this case. As a discussion of the proof of concept, the proposed photonic hook switch could be fabricated using the technologies of plano-convex-microsphere (PCM) lens laser nano-marking and two-photon polymerization (TPP) 3D printing [25, 26]. PCM lens is a dielectric lens made up of a high-index microsphere integrated with a plano-convex lens, which can deliver a high patterning resolution that is smaller than diffraction limit for a precision machining of the facet of the planar waveguide core in a trapezoid shape by a femtosecond (fs) laser [25]. TPP 3D printing is an additive fabrication technology to create micro/nano features relying on the photoresist solidification only occurring at the fs laser focus [26]. It can in-situ fill the micro/nanosized photonic wires of dielectric material layer by layer in the cavity created by PCM lens marking technology with a high degree of design freedom.

For the simulation setup in CST, the core material with the refractive index of $n = 1.44$ is used to encapsulate the trapezoid particle as long as the flat end facing the background medium of air ($n = 1$) with the open boundary condition along with $x$, $y$ and $z$ directions to simulate the planar waveguide transmission and photonic hook switching in the air. The incident plane wave is $y$-polarized and propagates along $z$ axis with the amplitude of 1 v/m. The trapezoid particle is placed as the prism encountering the light in $z$ direction. The short base, long base, height, thickness, and prism angle of the trapezoid are 2790 nm, 3720 nm, 2790 nm, 2790 nm, and 71.57° in this design, respectively. Also, the time domain solver and tetrahedral meshing with the size of 188 nm ($\lambda/7$) are used in the model to refine the simulation accuracy. The modelling diagram is shown in Figure 1 (b). The data of the Poynting vector in the modelling space generated by CST is exported and then independently processed by a MATLAB® programme to plot the contour of the Poynting vector magnitudes in a certain plane and a full 3D distribution of the Poynting vector streamlines entering the particle.

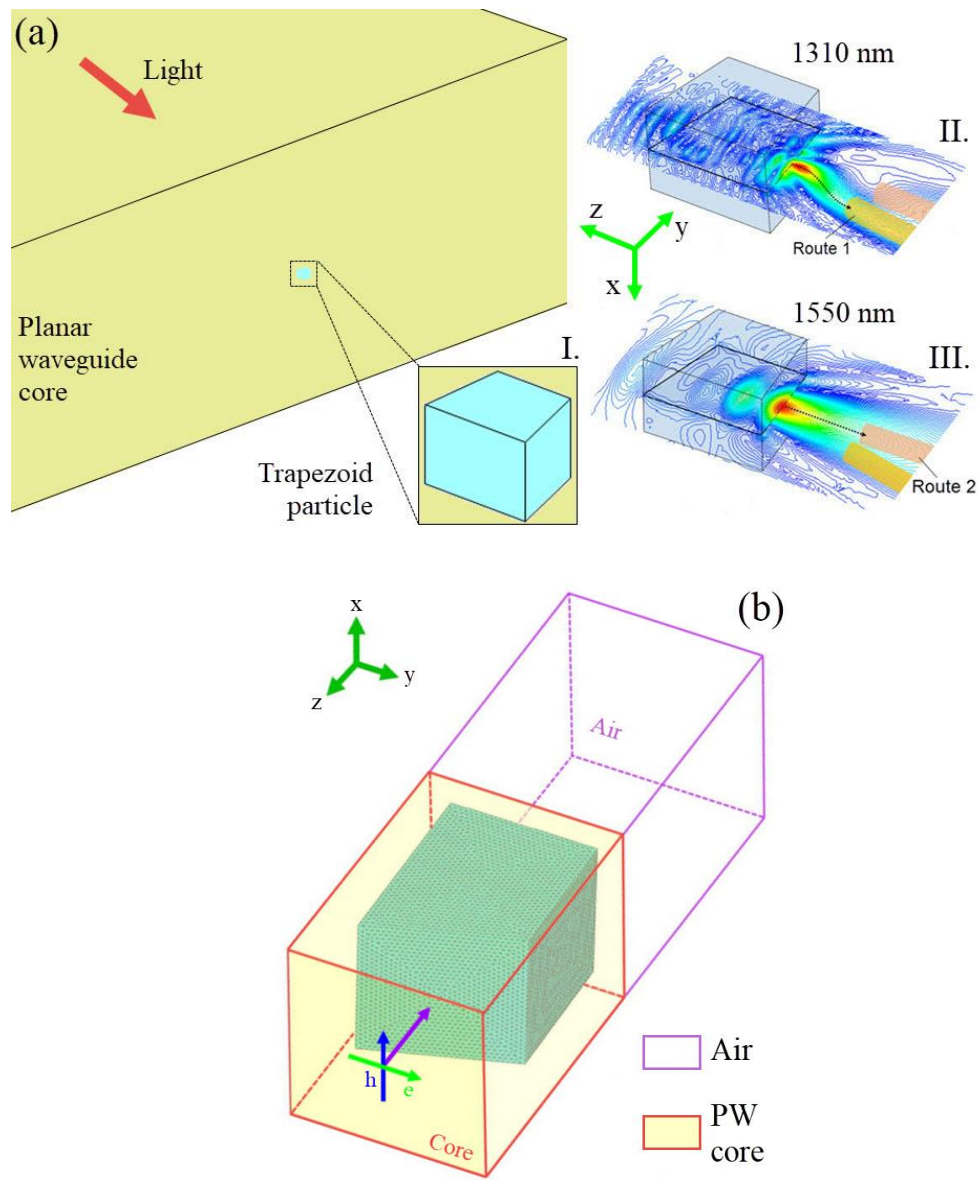

**Figure 1** (a) I. Diagram of photonic switch design; II & III. Principle of the proposed photonic switching at two different wavelengths (b) CST modelling diagram

### 2.2 Switching and 2D distributions
The dielectric particle in the proposed photonic switch design is individually irradiated by the plane-wave lights (optical signals) at the wavelengths of 1310 nm and 1550 nm, and the corresponding distributions of the squared electric field ($|E|^2$, $E$ is electric field) intensities and Poynting vectors ($S =$

$E \times H$, $H$ is magnetic field, unit: V·A/m$^2$) [27] are shown in Figure 2. The distributions of $|E|^2$ field in Figure 2 (a) and (b) demonstrate a clear separation of switching paths. In Figure 2 (a), the wavefront of 1310 nm light develops in a route approaching the side of the trapezoid long base, which forms a downward photonic hook. By contrast, the development of the photonic hook at the wavelength of 1550 nm is different from that for 1310-nm-wavelength light. It propagates in an almost straight path with a slightly upward curvature in the middle. The optical signals with the same wavelengths should be switched to the routes in accordance with the photonic hook curvatures. Theoretically, two ports can be set up in these routes to separately receive the corresponding signals.

Figure 2 (c) and (d) demonstrate the 2D Poynting vector distributions on the median section for the models irradiated by the optical signals with the wavelengths of 1310 nm and 1550 nm, respectively. The asymmetric vortexes of Poynting vector streamlines are shown in the dielectric particles in both figures. It is known that the photonic energy is able to flow in and out through these vortexes representing the singularities of Poynting vectors [28, 29]. Consequently, the layout of these vortexes can influence the flow direction and intensity of Poynting vectors, especially in the areas which are close to the particle boundaries. In Figure 2 (c) for 1310 nm optical signal, the prism of the dielectric particle leads the difference of phase velocity in the particle and generates 4 vortexes at the output end of the particle marked as v1, v2, v3 and v4 in the figure. Referring to the contour and flows of Poynting vectors in Figure 2 (c), it is shown that two saddle points – outward spiral v1 and inward spiral v2 are in high-intensity and low-intensity areas, respectively. This means that the photonic power flow passes into the median plane at v1 position in the upper half of the particle, meanwhile exiting the plane at v2 position in the lower half. The corresponding power flow exchange and influences of vortexes v3 and v4 around the output area of the particle make the flows of Poynting vectors in the generated photonic hook downward in general. However, it is noted that the photonic hook of 1310-nm-wavelength irradiation is divided into two high-magnitude areas (marked as a1 and a2) by a Poynting vector flow represented by a pink arrow in Figure 2 (c). This separation is unusual and rare in the discovered photonic hooks and different from the photonic hook in Figure 2 (a) for the squared electric field distribution. Rather than the downward photonic hook for the optical signal of 1310 nm wavelength, a slightly upward photonic hook is created by the same trapezoid particle for the incidence of 1550-nm-wavelength light. Two outward vortexes – v5 and v6 in Figure 2 (d) shape the flow distribution and produce two narrow low-magnitude regions compressing the high-magnitude area of the photonic hook in the middle.

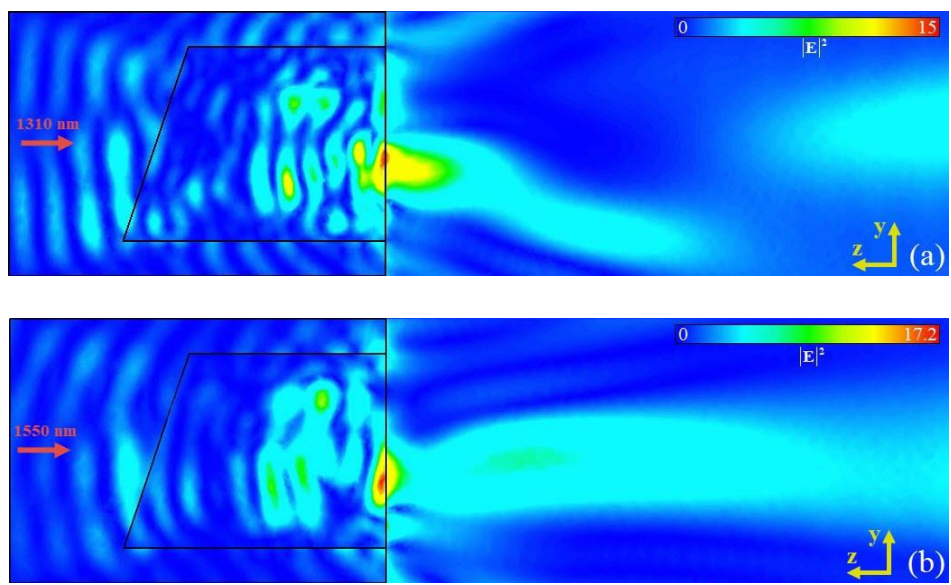

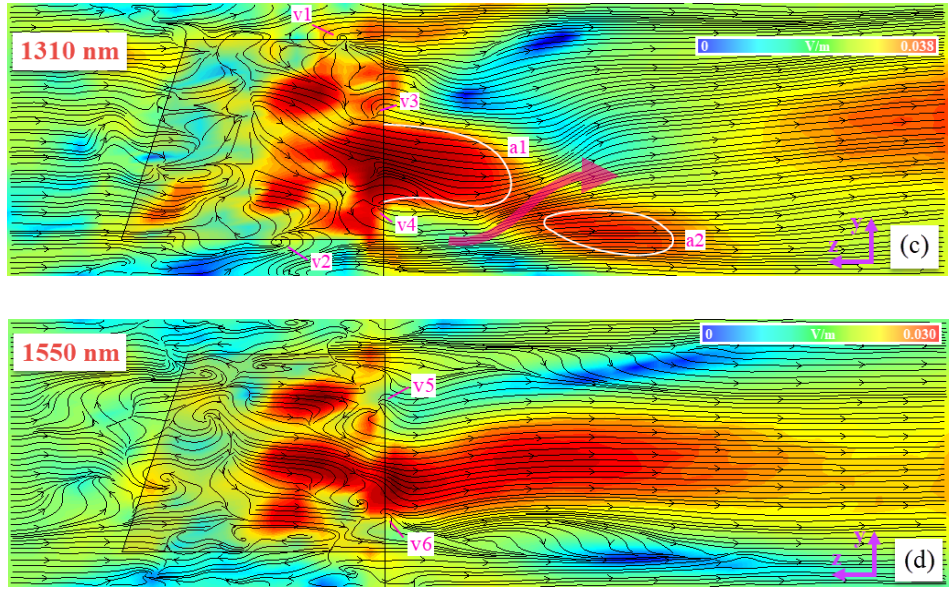

**Figure 2** The distributions of the squared electric field ($|E|^2$) for the irradiations with the wavelengths of 1310 nm (a) and 1550 nm (b). 2D Poynting vector distributions for the irradiations with 1310 nm (c) and 1550 nm (d).

### 2.3 3D Poynting vector analysis

The mechanism behind the switching of the optical signals at the wavelengths of 1310 nm and 1550 nm and their field distributions in Figure 2 is investigated by the 3D Poynting vector visualisation analysis [19]. All Poynting vectors entering the dielectric particle from the prism end are tracked and then visualised by an in-house developed MATLAB® programme. Because of a large amount of Poynting vector data imported in the programme, the plotted Poynting vector streamlines are marked in different colours to indicate the regions where they are tracked in the model. The colour codes are distributed along the *y* axis to equally divide the whole modelling space into three sections like the blue lines for the long-base section (-*y* direction), red lines for the middle section (around 0 position on the *y* axis ), and green lines for the short-base section (+*y* direction). An overview of 3D Poynting vector distribution for the model of 1310-nm-wavelength irradiation is shown in Figure 3 (a). All Poynting vector streamlines flow upwards from the prism end to the flat facet of the particle (vector arrows are hidden due to a high density of streamlines), however, the presented features of 3D distribution are significantly different from those in 2D format shown in Figure 2. The Poynting vector streamlines can continuously and freely develop in multiple directions and finally create a tangled-roots-like structure. Figure 3 (b) as a close-up view of the same model demonstrates the locations of the vortexes with the insets of two important ones - v1 and v2 indicated in Figure 2 as well. Except for the different spin directions of v1 and v2 shown in Figure 2, 3D visualisation of Poynting vectors can graphically represent the orientations of v1 and v2 and fully indicate this difference in Figure 3 (b). It is shown that the centre and spin-orbit of v1 is approximately perpendicular to the *z* axis and faces upward in -*z* direction, meanwhile, the same features are generally in the direction of 45° to the *x* axis for v2. In addition, the inner and outer circles of v1 are in red and green respectively, which means the photonic energy can flow to the middle and upper parts of the particle and modelling space through this vortex using multiple routes.

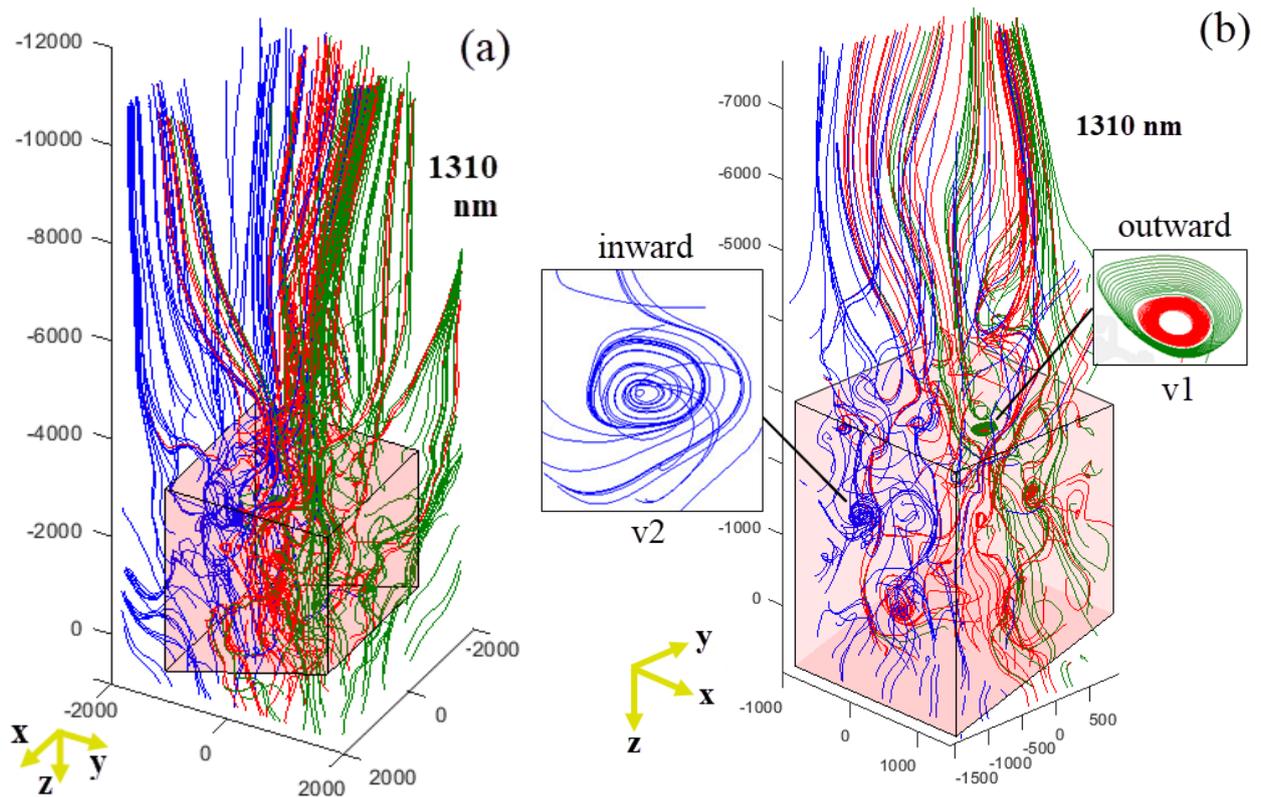

**Figure 3** Overview (a) and close-up view (b) of 3D Poynting vector distribution for the model of 1310 nm irradiation

The proposed switch enables the selection of the propagation routes for the optical signals at the wavelengths of 1310 nm and 1550 nm as shown in Figure 2 because of the opposite bending directions of photonic hooks. This phenomenon is unusual for two electromagnetic irradiations with such a small wavelength gap in the field of photonic hook study. This is mainly attributed to the separation of two high-magnitude areas of Poynting vectors as a1 and a2 shown in Figure 2 (c) for the irradiation at the wavelength of 1310 nm. The corresponding contour seems not perfectly matching with the flow direction and density of Poynting vectors in 2D Poynting vector distribution (Figure 2 (c)), especially for the gap between a1 and a2. Nevertheless, it can be properly explained by 3D Poynting vector analysis in this case. Here we use a 2D Poynting vector distribution of the median $yz$ plane ($x = 0$) in greyscale as a screen to divide the 3D Poynting vector visualisation model for the 1310-nm-wavelength irradiation into $+x$ segment and $-x$ segment, as the diagram shown in Figure 4 (a). View 1 ($+x$ segment visible) and view 2 ($-x$ segment visible) are defined as two observation angles to separately represent two segments of 3D Poynting vector distributions without allowing perspective in such a setup, as shown in Figure 4 (b) and (c). In these two figures, the profiles of the downward photonic hook and the high-magnitude areas – a1 and a2 are portrayed by black solid lines and yellow dashed lines, respectively. It is found that the profiles of a1 and a2 high-magnitude areas can be sequentially matched by the dense Poynting vector streamlines in the regions where are close to the median $yz$ plane of $x = 0$ in $+x$ segment (view 1) and $-x$ segment (view 2), respectively. The gap between a1 and a2 shown in Figure 2 (c) is attributed to this collective effect, and it leads to the unique downward photonic hook jointed by two high-magnitude areas for the irradiation at the wavelength of 1310 nm and spatially ensures optical isolation with the photonic hook (switching route) of 1550-nm-wavelength signal.

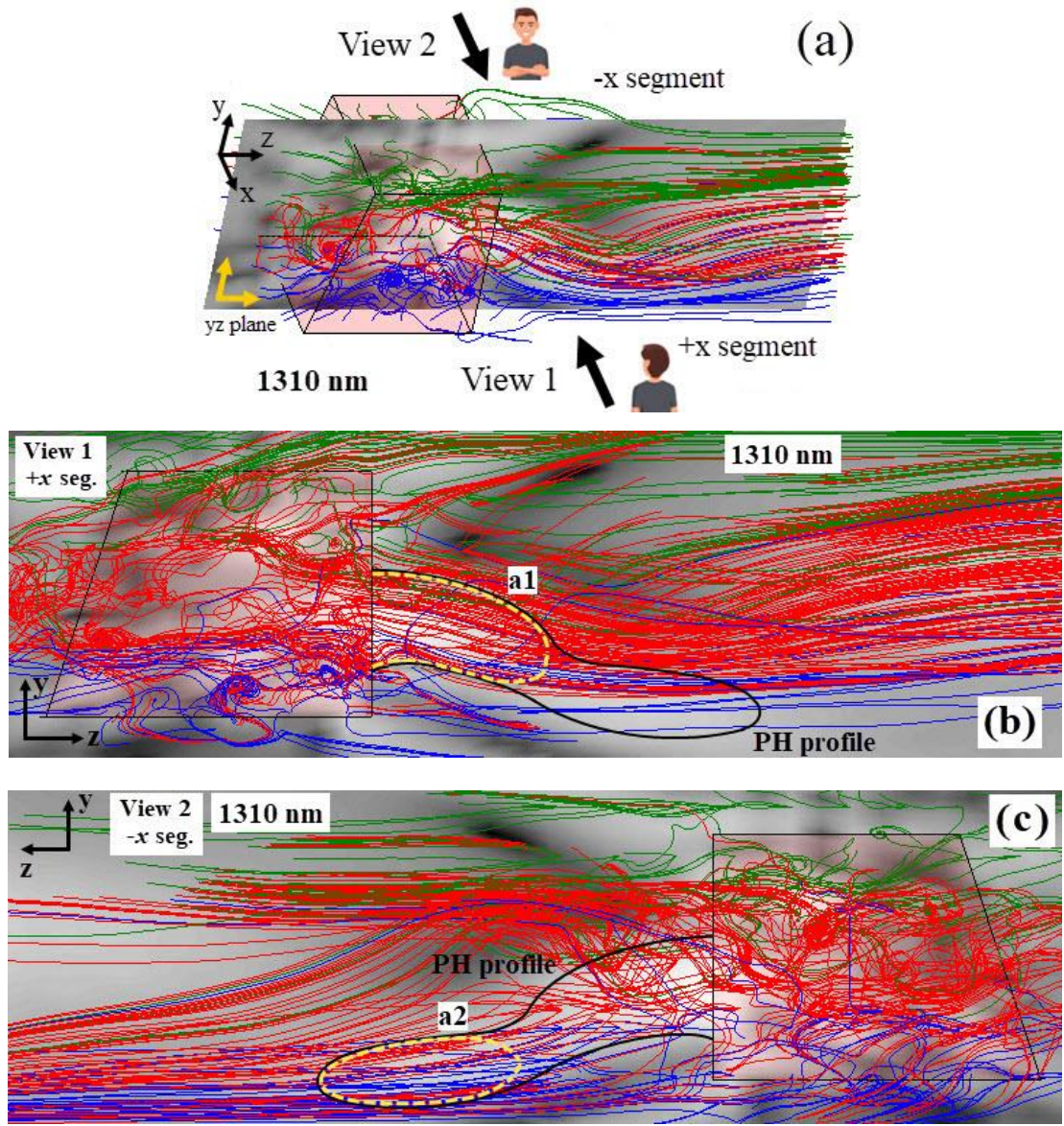

**Figure 4** (a) Diagram of view 1 and view 2 segments and 3D Poynting vector distributions for (b) view 1 +x segment and (c) view 2 -x segment of the model with the irradiation at the wavelength of 1310 nm. Orientations of the particle in (b) and (c) are reversed due to the change of frames of references for the observers in view 1 and view 2 as shown in (a).

Figure 5 (a) and (b) illustrate the overview and close-up views of 3D Poynting vector distribution for the model of 1550-nm-wavelength irradiation. Compared to the distributions for the model of 1310-nm-wavelength irradiation shown in Figure 3, the Poynting vector streamlines in Figure 5 are less swirling in the trapezoid particle and more straight in the volume of the generated photonic hook, which reflects a smaller number of vortexes shown in Figure 5. Two key vortexes – v5 and v6 are indicated in Figure 5 too. The photonic hook of the irradiation at a longer wavelength of 1550 nm appears in the region with the maximum density of Poynting vector streamlines, basically filling the central volume of the modelling space. These features are in accordance with the 2D Poynting distribution on the median *yz* plane for the same model shown in Figure 2 (d), which induces an almost straight, slightly upward photonic hook. Therefore, it is noted that the difference of spatial location of Poynting vector vortexes crucially influence the bending direction of the generated photonic hook and vary the subsequent propagation routes for the lights (optical signals) with similar wavelengths. This effect can be only represented by 3D Poynting vector distributions and cause multiple localised high-magnitude regions of Poynting vectors to constitute a complete single photonic hook.

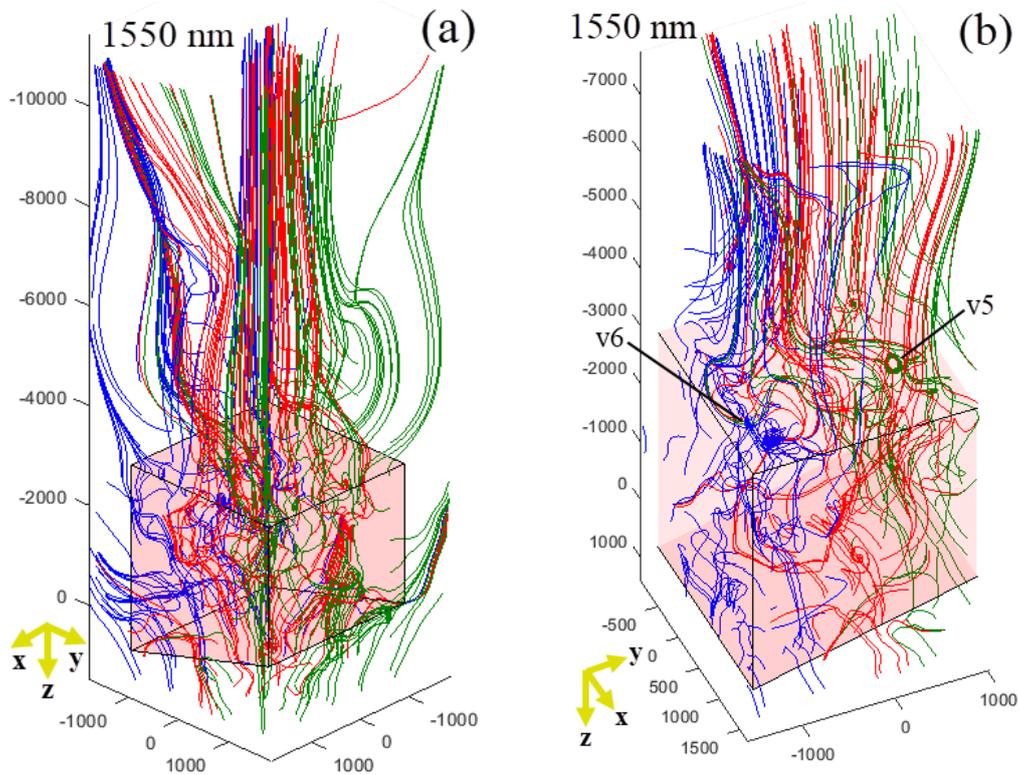

**Figure 5** Overview (a) and close-up view (b) of 3D Poynting vector distribution for the model of 1550 nm irradiation

**Conclusions**
The numerical simulation proves that the proposed near-field light-bending photonic switch as a Janus right-trapezoid-shaped dielectric particle embedded in the core facet of a planar waveguide can successfully switch and guide the optical signals at two common communication wavelengths – 1310 nm and 1550 nm to different propagation routes based on photonic hook phenomenon. The opposite bending directions are found for the generated photonic hooks, and the downward one for 1310-nm-wavelength irradiation is unusually separated into two high-magnitude areas by a Poynting vector flow. 3D Poynting vector analysis demonstrates that the corresponding high-magnitude areas can be sequentially matched by the Poynting vector streamlines in segments, which means that photonic hook can exist in a similar joining form. The corresponding mechanism is reported for the first time.

**Acknowledgements**
The authors received financial supports from the Centre for Photonic Expertise (CPE) with grant no. 81400 by Welsh European Funding Office (WEFO), the Royal Society International Exchanges Cost Share Scheme (reference no. IEC\R2\202040) UK, and the Russian Foundation for Basic Research (20-57-S52001). Part of the work was supported by the Tomsk Polytechnic University Development Program.

**References**
[1] R. C. Jaeger, *Microelectronic Circuit Design*, McGraw-Hill, New York, NY **1997**.
[2] M. M. Rao, *Optical communication*, Universities Press, Hyderabad, India, **2001**.
[3] J. M. Senior, M. Y. Jamro, *Optical fiber communications*, Pearson/Prentice Hall, Hoboken, NJ **2009**.
[4] C.Y. Jin, O. Wada, *J. Phys. D: Appl. Phys*. **2014**, 47, 133001.

[5] F. Testa, L. Pavesi, *Optical switching in next generation data centers*, Springer, Beilin **2017**.
[6] G. M. Bernstein, Y. Lee, A. Galver, J. Martensson, *IEEE/OSA J. Opt. Commun. Netw.* **2009**, 1, 1, 187–195.
[7] G. Cancellieri, *Single-mode optical fibres*, Elsevier Science, Amsterdam **2014**.
[8] D. Ozcelik, J. W. Parks, T. A. Wall, M. A. Stott, H. Cai, J. W. Parks, A. R. Hawkins, H. Schmidt, *Proc. Natl. Acad. Sci. U.S.A.* **2015**, 112, 12933–12937.
[9] D. Sadot, E. Boimovich, *IEEE Commun. Mag.* **1998**, 36, 50-55.
[10] S. Ramo, J. R. Whinnery, T. van Duzer, *Fields and waves in communications electronics*, John Wiley and Sons, NY **1984**.
[11] V. Prajzler, P. Nekvindova, O. Lyutakov, Radioengineering **2014**, 23, 3 776-782.
[12] J. N. Polky, G. L. Mitchell, *J. Opt. Soc. Am.* **1974**, 64, 274-279.
[13] I. V. Minin, O. V. Minin, *Diffractive Optics and Nanophotonics: Resolution below the Diffraction Limit*, Springer, Berlin **2016**.
[14] L. Yue, O. V. Minin, Z. Wang, J. N. Monks, A. S. Snalin, I. V. Minin, *Opt. Lett.* **2018**, 43, 771-774.
[15] I. V. Minin, O. V. Minin, G. M. Katyba, N. V. Chernomyrdin, V. N. Kurlov, K. I. Zaytsev, L. Yue, Z. Wang, D. N. Christodoulides, *Appl. Phys. Lett.* **2019**, 114, 031105.
[16] I. V. Minin, O. V. Minin, I. A. Glinskiy, R. A. Khalilubin, R. Malureanu, A. Lavrinenko, D. I. Yakubovsky, V. S. Volcow, D. S. Ponomarev, *Appl. Phys. Lett.* **2021**, 118, 131107.
[17] Y. E. Geints, O. V. Minin, L. Yue, I. V. Minin, *Ann. Phys. (Berlin)* **2021**, 2100192.
[18] C. M. Soukoulis, M. Wegener, *Nat. Photonics* **2011**, 5, 523–530.
[19] L. Yue, B. Yan, J. N. Monks, R. Dhama, C. Jiang, O. V. Minin, I. V. Minin, Z. Wang, *Sci. Rep.* **2019**, 9, 1-8.
[20] S. Y. Lee, T. Y. Jeong, S. Jung, K. J. Yee, *Phys. Status Solidi B* **2018**, 256, 1800417.
[21] C. Z. Tan, *J. Non-Cryst. Solids* **1998**, 223, 158-163.
[22] MICROPOSIT® S1800® photoresist datasheet https://amolf.nl/wp-content/uploads/2016/09/datasheets_S1800.pdf
[23] J. M. Elson, *Opt. Express* **2001**, 9, 461-475.
[24] J. E. G´omez-Correa, S. E. Balderas-Mata, A. Garza-Rivera, A. Jaimes-N´ajera, J. P. Trevino, V. Coello, J. Rogel-Salazar, and S. Ch´avez-Cerdac, *Revista Mexicana de F´ısica E* **2019**, 65 218-222.
[25] B. Yan, L. Yue, J. N. Monks, X. Yang, D. Xiong, C. Jiang, Z. Wang, *Opt. Lett.* **2020**, 45, 1168-1171.
[26] PI. Dietrich, M. Blaicher, I. Reuter, M. Billah, T. Hoose, A. Hofmann, C. Caer, R. Dangel, B. Offrein, U. Troppenz, M. Moehrle, W. Freude, C. Koos, *Nature Photon.* **2018**, 12, 241–247.
[27] J. A. Stratton, *Electromagnetic Theory (1st ed.)*, McGraw-Hill, NY **1941**.
[28] C. F. Bohren, D. R. Huffman, *Absorption and scattering of light by small particles*. Wiley, NY **1983**.
[29] Z. B. Wang, B. S. Luk'yanchuk, M. H. Hong, Y. Lin, T. C. Chong, *Phys. Rev. B* **2004**, 70, 035418.